\documentclass[aps,prl,twocolumn,superscriptaddress,showpacs,floatfix]{revtex4-1}
\usepackage{amsmath,amssymb,amsfonts,float,graphics,epsfig,epstopdf,color,verbatim,tabularx,bm,multirow}
\usepackage{hyperref}
\hypersetup{
colorlinks = true,
linkcolor = [rgb]{0.70,0.13,0.13},
citecolor = [rgb]{0.13,0.55,0.13},
urlcolor = [rgb]{0.25,0.41,0.88}}

\newcommand{\dd}{\mathrm{d}}
\newcommand{\ii}{\mathrm{i}}
\newcommand{\SU}{\mathrm{SU}}
\newcommand{\U}{\mathrm{U}}
\renewcommand{\O}{\mathrm{O}}
\newcommand{\SO}{\mathrm{SO}}

\newcommand{\dsZ}{\mathbb{Z}}

\newcommand{\scL}{{\mathcal{L}}}

\DeclareSymbolFont{sfletters}{OML}{cmbrm}{m}{it}
\DeclareMathSymbol{\sfeps}{\mathord}{sfletters}{"22}

\newcommand{\vect}[1]{{\bm{#1}}}

\newcommand{\eq}[1]{\begin{equation}#1\end{equation}}
\newcommand{\eqs}[1]{\begin{equation}\begin{split}#1\end{split}\end{equation}}
\newcommand{\eqnref}[1]{Eq.\,\eqref{#1}}
\newcommand{\figref}[1]{Fig.\,\ref{#1}}
\newcommand{\tabref}[1]{Tab.\,\ref{#1}}

\graphicspath{{Figures/}}

\begin{document}

\title{Emmy Noether looks at the deconfined quantum critical point}

\author{Nvsen Ma}
\affiliation{Beijing National Laboratory for Condensed Matter Physics and Institute of Physics, Chinese Academy of Sciences, Beijing 100190, China}
\author{Yi-Zhuang You}
\affiliation{Department of Physics, University of California, San Diego, California 92093, USA}
\affiliation{Department of Physics, Harvard University, Cambridge, Massachusetts 02138, USA}
\author{Zi Yang Meng}
\affiliation{Beijing National Laboratory for Condensed Matter Physics and Institute of Physics, Chinese Academy of Sciences, Beijing 100190, China}
\affiliation{Department of Physics, The University of Hong Kong, Pokfulam Road, Hong Kong, China}
\affiliation{CAS Center of Excellence in Topological Quantum Computation and School of Physical Sciences, University of Chinese Academy of Sciences, Beijing 100190, China}
\affiliation{Songshan Lake Materials Laboratory, Dongguan, Guangdong 523808, China}

\date{\today}

\begin{abstract}
Noether's theorem is one of the fundamental laws of physics, relating continuous symmetries and conserved currents. Here we explore the role of Noether's theorem at the deconfined quantum critical point (DQCP), which is a quantum phase transition beyond the Landau-Ginzburg-Wilson paradigm. It was expected that a larger continuous symmetry could emerge at the DQCP, which, if true, should lead to conserved current at low energy. By identifying the emergent current fluctuation in the spin excitation spectra, we can quantitatively study the current-current correlation in large-scale quantum Monte Carlo simulations. Our results reveal the conservation of the emergent current, as signified by the vanishing anomalous dimension of the current operator, and hence provide supporting evidence for the emergent symmetry at the DQCP. Our study demonstrates an elegant yet practical approach to detect emergent symmetry by probing the spin excitation, which could potentially guide the ongoing experimental search for DQCP in quantum magnets.
\end{abstract}

\maketitle

{\it{Introduction.-}} Noether's theorem is a profound theorem in physics that states every continuous (differentiable) symmetry of a physical system is associated with a corresponding conservation law~\cite{Noether1918}. Well-known examples include momentum and energy conservations, when the system respects space and time translation symmetries. The conservation law usually manifests itself in the form of a conserved current $J_\mu$, which satisfies the equation $\partial^\mu J_\mu=0$. Likewise, the observation of a conserved current in a physical system usually serves as a direct evidence of the associated continuous symmetry.

In this paper, we introduce an explicit application of the Noether's theorem in identifying the emergent continuous symmetry in an exotic quantum phase transition -- the deconfined quantum critical point (DQCP)~\cite{ashvinlesik,deconfine1,deconfine2,Sandvik2007,YQQin2017}. The DQCP describes a direct continuous transition between two phases that spontaneously breaks very different symmetries. In particular, we focus on a type of DQCP which is only recently identified in quantum Monte Carlo (QMC) simulations~\cite{YQQin2017,Ma2018a}, dubbed the easy-plane DQCP. It is a direct quantum phase transition in a (2+1)D quantum spin model between the antiferromagnetic XY (AFXY) ordered phase and the columnar valence bond solid (VBS) phase. The AFXY (VBS) phase is described by the ordering of a two-component spin order parameter $(N_x,N_y)$ (dimer order parameter $(D_x,D_y)$), which spontaneously breaks the in-plane $\U(1)$ spin rotation symmetry ($\dsZ_4$ lattice rotation symmetry). At the transition point, both the spin and dimer order parameters fluctuates strongly but with vanishing expectation values, such that the microscopic $\U(1)\times\dsZ_4$ is restored. Remarkably, it has been suggested that the low-energy critical fluctuations could respect an even larger emergent $\O(4)$ symmetry, which corresponds to the full rotation of the combined four-component order parameter $(D_x,D_y,N_x,N_y)$. The emergent $\O(4)$ symmetry, if exists, is a hallmark of the easy-plane DQCP~\cite{SO5,maxryan17,Senthil2018}.

\begin{figure}[htp!]
\begin{center}
\includegraphics[width=0.8\columnwidth]{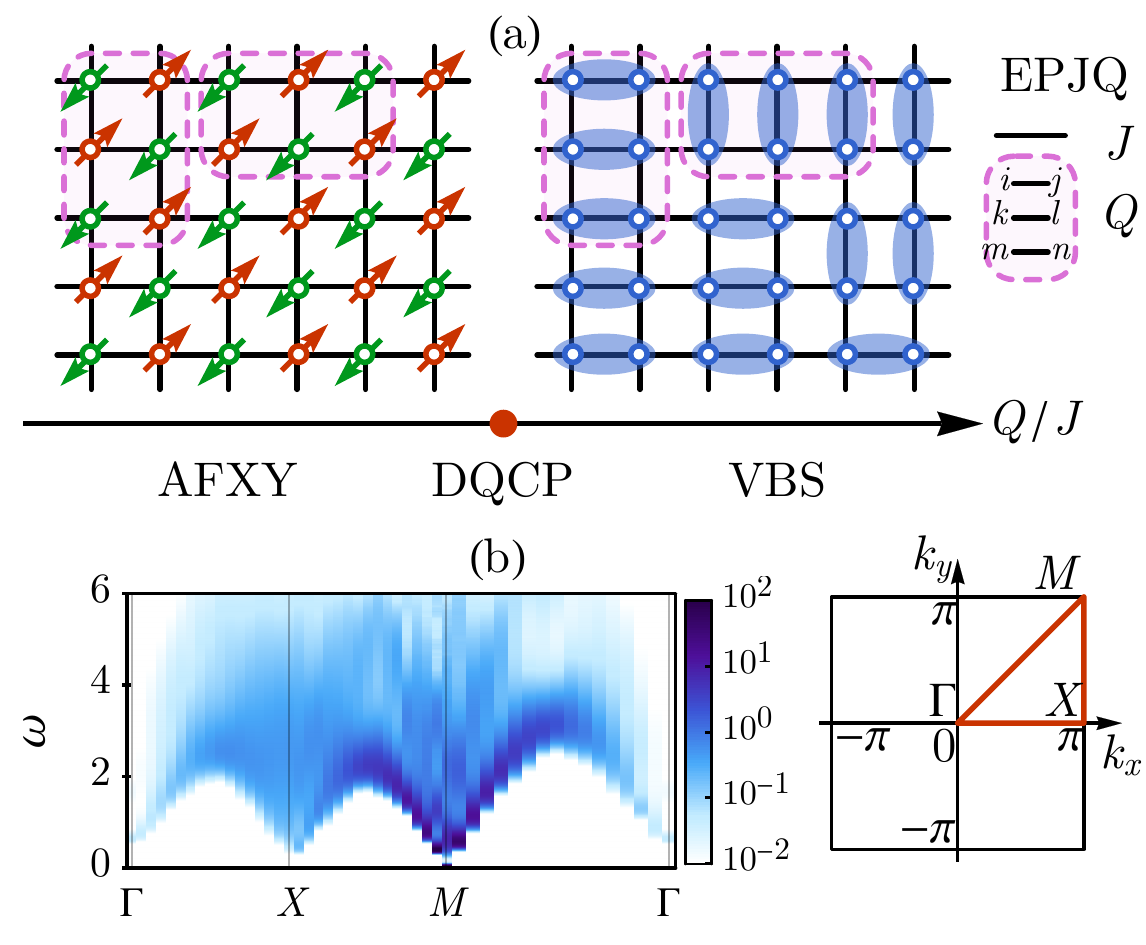}
\caption{(a) The easy-plane J-Q (EPJQ) model and its phase diagram. The $Q$ term describes the three-dimer interaction in both horizontal and vertical directions, with the arrangement of site indices shown on the right. (b) Obtained spin excitation spectra in $S^{x}$ channel of the EPJQ model at the DQCP in our previous work~\cite{Ma2018a}. Darker color indicates higher intensity. The high symmetry points in the Brillouin zone (BZ) are defined on the right.}
\label{fig:fig1}
\end{center}
\end{figure}

Several methods have been developed to test the emergent symmetry at the DQCP, including comparing the critical exponents in spin and VBS channels~\cite{Lou2009,Block2013,Shao2016,YQQin2017,YuhaiLiu2018,Sandvik2007,Kaul12,Nahum2015a,Nahum2015b,Jonathan2017,Ma2018a}, plotting the order parameter histograms~\cite{Sandvik2007,Lou2009,Nahum2015b,XFZhang2017,BWZhao2018,Sato2017,YCWang2017,XYXu2018}, and probing the degenerated correlation spectra of spin-0 and spin-1 excitations~\cite{Wang:2018Cr,JYLee2019}. But the conserved Noether current that is directly associated with the emergent continuous symmetry~\cite{Noether1918,hermele2005} has not been measured (with a recent work in 1D system with emergent U(1) symmetries detected by conserved currents~\cite{RZHuang2019}). In this work, we directly probe the $\SO(4)$ current fluctuation at the easy-plane DQCP and verify the current conservation by measuring the scaling dimensions of the current-current correlations.

Based on the field theoretical analysis, which captures the emergent $\O(4)$ symmetry of the DQCP, we are able to identify different components of the $\SO(4)$ current operator with the microscopic spin operator at different momenta. We then measure imaginary-time correlation functions of these spin fluctuations at the designated momenta in large-scale QMC simulations to determine the critical scaling of the current operator. The conservation of the Noether current $\partial^\mu J_\mu=0$ implies that the flux of the current $\int_{\partial\Omega} \epsilon^{\mu\nu\lambda}J_\mu\dd x_\nu\dd x_\lambda$ through a close manifold $\partial\Omega$ in the (2+1)D spacetime must remain constant. As a result, the conserved current should follow a precise scaling law $J_\mu\sim x^{-2}$ with the scaling dimension pinned at 2, which is an integer instead of a usually fractional critical exponent for generic operators~\footnote{We note that at there are cases for CFT where the leading power laws of three-point (and higher) correlations are not given simply by the scaling dimensions of the lattice operators involved, but complications can happen such that faster decay is observed due to exact cancellations of contributions from the fields and currents under conformal symmetry. This is explicitly shown in 1D case in by Pranay et al. in Ref.~\cite{Pranay2018}}. By a systematic finite-size-scaling of the numerical data, we are able to make quantitative comparison of the measured scaling dimensions with the field theoretical expectation.

Our results are in strong support of the emergent $\O(4)$ symmetry, up to the largest system size $L=96$. The emergent $\O(4)$ symmetry generally points to a continuous transition at the easy-plane DQCP, although we can not rule out the possibility of a symmetry-enhanced first-order transition (plausibly due to a pseudocriticality\cite{SO5}). Such possibility was recently reported by Zhao, Weinberg and Sandvik~\cite{BWZhao2018} and  Serna and Nahum~\cite{Serna2018} in several related models.
Our attempt of bridging the conserved current correlation from basic law of physics with large-scale numerical calculation of quantum many-body systems provides a complementary and elegant way of identifying the emergent continuous symmetry at quantum phase transitions (not necessarily to be continuous), which are becoming ubiquitously present in the new paradigms of quantum matter such as various forms of DQCP~\cite{ashvinlesik,deconfine1,deconfine2,Sandvik2007,Kaul2013a,Nahum2015b,XFZhang2017,YQQin2017,Sato2017,BWZhao2018,Serna2018,YuhaiLiu2018}, frustrated magnets~\cite{Moessner2001,Isakov2003,YCWang2017,XiaoYanXu2018}, interacting topological phases~\cite{Assaad2016,Gazit2018,YCWang2018,YuhaiLiu2018} and quantum electrodynamic systems~\cite{Karthik2017Fl,XiaoYanXu2018,Senthil2018}. Our numerical study of the conserved current at an emergent $\O(4)$ symmetric DQCP will also guide future spectroscopy experiments, neutron scattering for example, in search of DQCP in candidate materials such as the Shastry-Sutherland lattice compond $\mathrm{SrCu_2 (BO_3)_2}$~\cite{JGuo2019,JYLee2019}.

{\it{Model.-}} The easy-plane DQCP was reported in our previous QMC simulation of the easy-plane J-Q (EPJQ) model~\cite{YQQin2017}, which is a spin model on a two-dimensional square lattice, as illustrated in \figref{fig:fig1}(a). The model is described by the following Hamiltonian
\begin{equation}
\label{eq:EPJQ}
H_\text{EPJQ}=-J\sum_{\langle ij\rangle}(P_{ij}+\Delta S_i^zS_j^z)-Q\hskip-2mm\sum_{\langle ijklmn \rangle}\hskip-2mm
P_{ij}P_{kl}P_{mn},
\end{equation}
where $\vect{S}_i=(S_i^x,S_i^y,S_i^z)$ denotes the spin-1/2 operator on each site $i$ and $P_{ij}=\tfrac{1}{4}-\boldsymbol{S}_i\cdot\boldsymbol{S}_j$ is the spin-singlet projection operator across the bond $\langle ij\rangle$. The summation of $\langle ijklmn\rangle$ runs over all the six-site clusters containing three parallel bonds $\langle ij\rangle, \langle kl\rangle, \langle mn\rangle$, as shown in Fig.~\ref{fig:fig1}(a), which can be arranged either horizontally or vertically.

The $J$ term describes the nearest neighboring antiferromagnetic ($J>0$) spin interaction with an easy-plane anisotropy introduced by the $\Delta S_i^zS_j^z$ term ($0<\Delta\leq 1$). The $Q$ term describes the attractive ($Q>0$) interaction among three adjacent parallel dimers. The model admits sign-free QMC simulations for all range of parameters. At $\Delta=0$, the model goes back to the SU(2) symmetric J-Q$_3$ model\cite{Lou2009,Sen2010,Sandvik2010}. With finite $\Delta>0$, the SU(2) spin rotation symmetry is explicitly broken down to its U(1) subgroup, describing the in-plane rotation of XY spins. We define the tuning parameter $q=\frac{Q}{J+Q}$ (such that $0\leq q\leq 1$). When $q\to0$, the model favors the antiferromagnetic ordered phase of XY spins, denoted as the AFXY phase. When $q\to1$, the VBS phase is favored. The AFXY and VBS order parameters are defined as
\begin{equation}\label{eq:orders}
N_{x}=\sum_i (-)^{x_i+y_i} \langle S_i^{x}\rangle,\quad D_{x}=\sum_i (-)^{x_i}\langle P_{i,i+\hat{x}}\rangle,
\end{equation}
and $N_y,D_y$ are similarly defined under $x\leftrightarrow y$, where $\vect{r}_i=(x_i,y_i)$ labels the coordinate of site $i$ on the square lattice and $\hat{x}=(1,0), \hat{y}=(0,1)$ are lattice unit vectors. In Ref.~\onlinecite{YQQin2017} it is shown that at $\Delta=1/2$, the EPJQ model exhibits a direct quantum phase transition 
between the AFXY and VBS phases at $q_c=(\frac{Q}{J+Q})_c=0.62(1)$, realizing the easy-plane DQCP. As argued based on dualities~\cite{xudual,SO5,Senthil2018}, the critical point is expected to exhibit an emergent $\O(4)$ symmetry at low-energy, which rotates both the AFXY and VBS order parameters together as an $\O(4)$ vector $\vect{n}=(n^1,n^2,n^3,n^4)=(D_x,D_y,N_x,N_y)$. 

{\it{Conserved currents.-}} According to Noether's theorem, the proposed $\O(4)$ emergent symmetry at the easy-plane DQCP must be accompanied with the corresponding emergent conserved currents. The goal of this work is to test these emergent conserved currents in numerics. We are interested in the continuous $\SO(4)$ subgroup of $\O(4)$, which has six Lie group generators. Each generator $A^{ab}$ is labeled by a pair of ordered indices $a<b$ taken from $a,b=1,2,3,4$, which generates the rotation between $n^a$ and $n^b$ components of the $\O(4)$ vector $\vect{n}$. If the $\O(4)$ symmetry indeed emerges at low-energy, we should be able to observe six conserved currents $J_\mu^{ab}$, each corresponds to a generator $A^{ab}$ with three space-time components labeled by $\mu=0,1,2$ (the temporal component is the conserved charge density). By a simple counting, there are $6\times3=18$ components of the $\SO(4)$ conserved currents $J_\mu^{ab}$ (6 from $ab$ and 3 from $\mu$ indices). Their low-energy and long-wavelength fluctuations are expected to appear as  quantum critical fluctuations at the DQCP.

However we do not aim to observe all the 18 components of $J_\mu^{ab}$ here, instead, we will focus on those that are detectable in the \emph{spin} excitation spectrum, which are simpler to measure in QMC simulations (compared to the dimer excitations) and are more relevant to scattering experiments. It was found in Ref.~\onlinecite{Ma2018a} that at the critical point, the spin excitation spectrum will become gapless at four momentum points $\vect{Q}$: $(0,0)$, $(\pi,0)$, $(0,\pi)$ and $(\pi,\pi)$, as shown in Fig.\ref{fig:fig1}(b) for the $S^x$ channel (the $S^z$ channel shares a similar shape). We will label these low-energy fluctuations by
\begin{equation}
\vect{S}_\vect{Q}(q)=\sum_{i}\vect{S}_i(\tau) e^{\ii q_0\tau-\ii(\vect{Q}+\vect{q})\cdot\vect{r}_i},
\end{equation}
where $\vect{S}_i(\tau)$ is the spin operator on site $i$ at imaginary time $\tau$ and $q=(q_0,\vect{q})$ labels the imaginary frequency and momentum. It turns out that 5 of the 18 components of $J_\mu^{ab}$ appear in the spin excitation spectrum, as summarized in Tab.~\ref{tab:currents}. These identifications are made by resorting to the field theory description\cite{hermele2005,ranwen,You2018BTT} of the easy-plane DQCP. The detail derivations are provided in Sec.~I of Supplemental Material (SM)~\cite{suppl}. 

\begin{table}[htbp]
\caption{Identification between low-energy spin excitations and emergent $\SO(4)$ conserved currents}
\begin{center}
\begin{tabular}{cccc}
spin & $\vect{Q}$ & current  & related symmetry\\
\hline
$S^x$ & $(\pi,0)$ & $J_2^{23}$ & $(D_y,N_x)$ rotation\\
$S^x$ & $(0,\pi)$ & $J_1^{13}$ & $(D_x,N_x)$ rotation\\
$S^y$ & $(\pi,0)$ & $J_2^{24}$ & $(D_y,N_y)$ rotation\\
$S^y$ & $(0,\pi)$ & $J_1^{14}$ & $(D_x,N_y)$ rotation\\
$S^z$ & $(0,0)$ & $J_0^{34}$ & $(N_x,N_y)$ rotation\\
\end{tabular}
\end{center}
\label{tab:currents}
\end{table}

The first four currents $J_2^{23},J_1^{13},J_2^{24},J_1^{14}$ are associated to the emergent rotational symmetry between AFXY and  VBS order parameters~\cite{Ma2018a}. There is no such symmetry at the lattice level, as is evident from the distinct forms of the order parameters in \eqnref{eq:orders}. The emergent AFXY-VBS rotation symmetry is the most crucial part of the $\O(4)$ symmetry group that glues the microscopic $\U(1)$ and $\dsZ_4$ symmetries together. The observation of the conservation law for these currents will provide a direct evidence for the emergent $\O(4)$ symmetry. Given that the above four currents are related by the microscopic $\U(1)\times\dsZ_4$ symmetry, it is sufficient to only focus on $J_2^{23}$, which corresponds to the $S_{(\pi,0)}^x$ spin fluctuation. By measuring whether $S_{(\pi,0)}^x$ has a vanishing anomalous dimension, we can determine whether $J_2^{23}$ is conserved or not. For comparison, we also study the last (microscopic) conserved current $J_0^{34}$ in Tab.~\ref{tab:currents}, in association to the microscopic $\U(1)$ symmetry, which appears as the $S_{(0,0)}^z$ spin fluctuation.

{\it{Numerical Results.-}} Suppose the easy-plane DQCP has the proposed $\O(4)$ emergent symmetry, this will put a strong constraint on the correlation function of current operators and the consequence can be tested in our QMC simulation. For an emergent conserved current $J_{\mu}^{ab}$, its correlation function will be universally given by
\begin{equation}
\label{eq:JJcon}
\langle J_{\mu}^{ab}J_{\nu}^{ab}\rangle\sim|q|\Big(\delta_{\mu\nu}-\frac{q_{\mu}q_{\nu}}{|q|^2}\Big), (a,b=1,2,3,4),
\end{equation}
which will not receive corrections from gauge fluctuations and spinon interactions. 
Using the operator correspondence in \tabref{tab:currents}, the current-current correlation in the field theory can be translated to the spin-spin correlation in the lattice model as
\begin{equation}\label{eq:SSinq}
\begin{split}
\langle S_{(\pi,0)}^xS_{(\pi,0)}^x\rangle&\sim\langle J_{2}^{23}J_{2}^{23}\rangle\sim(q_0^2+q_1^2)/|q|^{1-\eta_{(\pi,0)}^x},\\
\langle S_{(0,0)}^zS_{(0,0)}^z\rangle&\sim\langle J_{0}^{34}J_{0}^{34}\rangle\sim(q_1^2+q_2^2)/|q|^{1-\eta_{(0,0)}^z}.
\end{split}
\end{equation}
We have introduced two anomalous exponents $\eta_{(\pi,0)}^x$ and $\eta_{(0,0)}^z$ for general considerations. They will vanish separately if their corresponding currents are indeed conserved. In particular, the vanishing $\eta_{(\pi,0)}^x$ will be non-trivial, as it corresponds to the conservation of the emergent current $J_2^{23}$ of AFXY-VBS rotation, which is not expected at the microscopic level. In this way, we can determine the emergent symmetry from the vanishing anomalous exponent of the Noether current. Only a single exponent is needed in this approach. This is different from measuring the non-vanishing anomalous exponents of the order parameters in previous works for O(4)~\cite{YQQin2017,XFZhang2017,Karthik2017Fl} and SO(5)~\cite{Sandvik2007,Lou2009,Sandvik2010,Kaul12,Shao2016} cases, where one needs to compare exponents of different order parameters to determine the emergent symmetry. The conserved current correlation offers an independent probe of emergent continuous symmetry, which is complementary to previous approaches such as the order parameter histogram~\cite{Sandvik2007,Nahum2015b,Sato2017}.

\begin{figure}[htp]
\includegraphics[width=0.8\columnwidth]{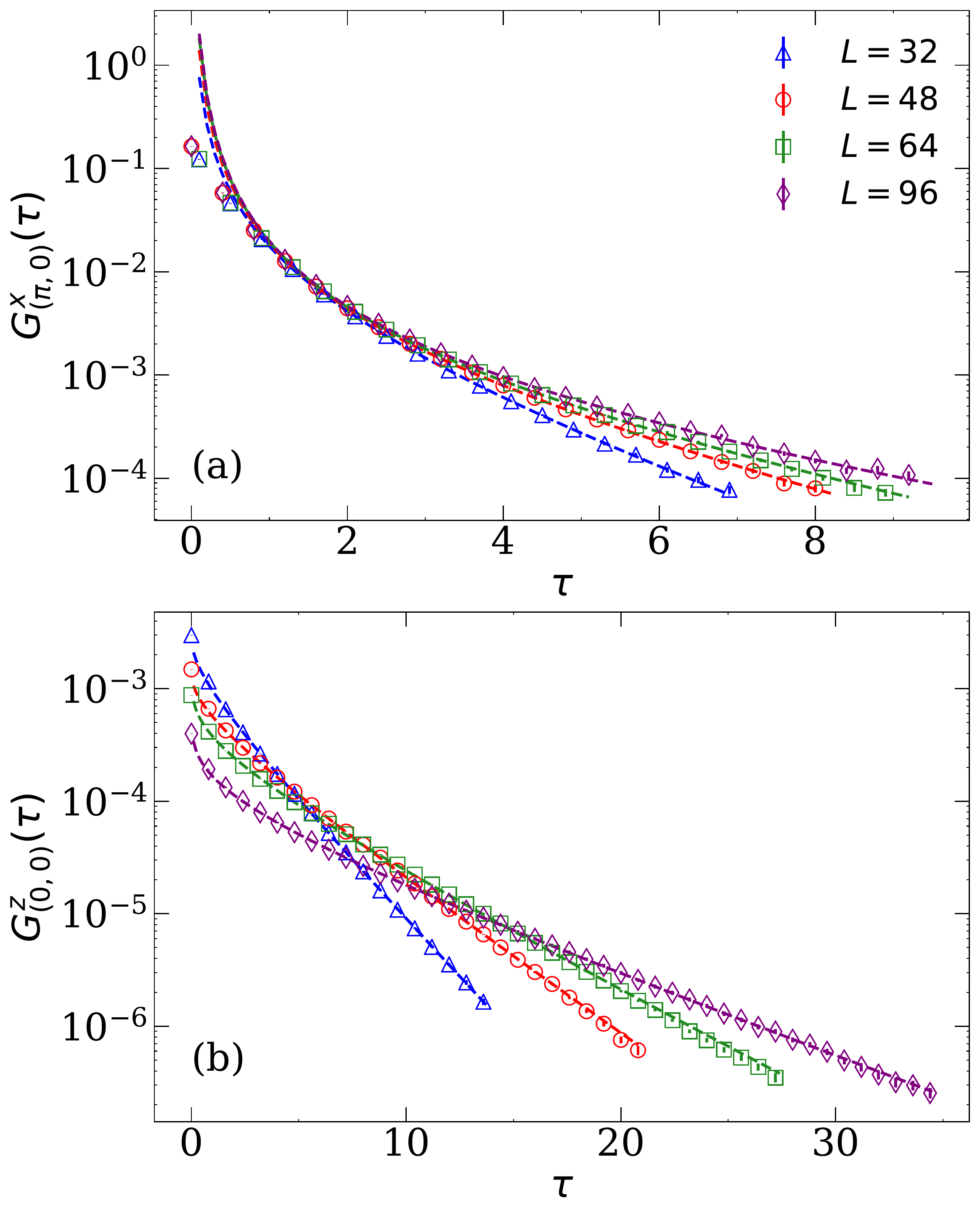}
\caption{Spin-spin correlation functions measured at the DQCP of model in Eq.~\eqref{eq:EPJQ}, $q_c=0.62$, with inverse temperature $\beta=2L$ and $L=32,48,64,96$. (a) $G^{x}_{(\pi,0)}(\tau,\vect{q})$ and (b) $G^{z}_{(0,0)}(\tau,\vect{q})$. All curves in the figures are fitting results using Eq.~\eqref{eq:bessel}.}
\label{fig:fig2}
\end{figure}

\begin{figure}[htp!]
\includegraphics[width=0.9\columnwidth]{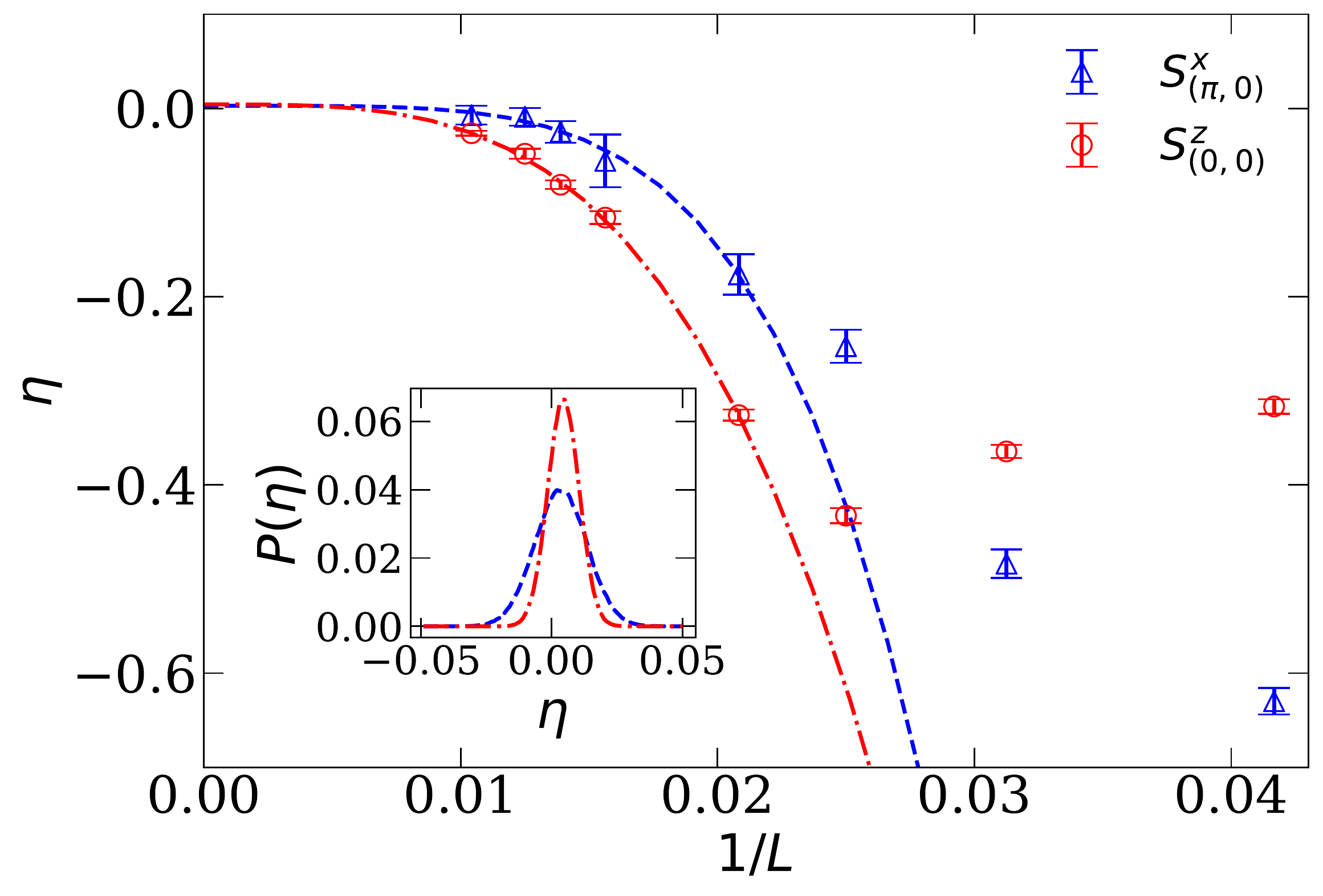}
\caption{The anomalous scaling dimensions $\eta^{z}_{(0,0)}$ and $\eta^{x}_{(\pi,0)}$ obtained from the fitting in Fig.\ref{fig:fig2}. As $L$ increases, both $\eta^{z}_{(0,0)}$ and $\eta^{x}_{(\pi,0)}$ extrapolate to 0, consistent with the prediction of the conserved currents in these two channels. Inset shows the histogram of the extrapolated $\eta$ obtained from many Gaussion noise realizations of the finite size $\eta$ values. The histogram of both $\eta^{z}_{(0,0)}$ and $\eta^{x}_{(\pi,0)}$ are centered at zero.}
\label{fig:fig3}
\end{figure}

In QMC simulations, the spin-spin correlation $G_\vect{Q}^a(\tau,\vect{q})\equiv\sum_{ij}\langle S^a_i(\tau)S^a_j(0)\rangle e^{\ii(\vect{Q}+\vect{q})\cdot(\vect{r}_i-\vect{r}_j)}$ (for $a=x,y,z$) can be directly measured in the imaginary time domain and the momentum space. In order to make comparison with the numerics, we need to Fourier transform the previous field theory predictions in \eqnref{eq:SSinq} from Matsubara frequency to imaginary time, following $G_{\vect{Q}}^{a}(\tau,\vect{q})=\int\dd q_0 e^{-\ii q_0 \tau} \langle S_{\vect{Q}}^a(-q)S_\vect{Q}^a(q)\rangle$. The results are
\begin{eqnarray}
G^x_{(\pi,0)}(\tau,\vect{q})&\propto & q_2^2F_{\frac{\eta}{2}}(\tau,\vect{q})+\tfrac{\eta+1}{2}F_{\frac{\eta}{2}+1}(\tau,\vect{q}))|_{\eta=\eta_{(\pi,0)}^x},\nonumber\\
G^z_{(0,0)}(\tau,\vect{q})&\propto & \vect{q}^2F_{\frac{\eta}{2}}(\tau,\vect{q})|_{\eta=\eta_{(0,0)}^z},
\label{eq:bessel}
\end{eqnarray}
where $F_\alpha(\tau,\vect{q})=|\frac{2\vect{q}}{\tau}|^\alpha K_\alpha(|\vect{q}\tau|)$ and $K_\alpha(|\vect{q}\tau|)$ is the $\alpha$-th order Bessel $K$-function (detailed derivations of Eq.~\eqref{eq:bessel} are given in Sec.II of SM~\cite{suppl}). The anomalous dimension $\eta_{(\pi,0)}^x$  and $\eta_{(0,0)}^z$ are fitting parameters to be determined from the data. The numerical determination of these exponents from finite size QMC results will be the focus of narrative below.

Figures ~\ref{fig:fig2} (a) and (b) depict the imaginary time correlations $G^{x}_{(\pi,0)}(\tau,\vect{q}=0)$ and $G^{z}_{(0,0)}(\tau,\vect{q})$ respectively. We note that around $\vect{Q}=(\pi,0)$, the spin-spin correlation remains finite, so we take the QMC measurements at $(\pi,0)$; whereas around $\vect{Q}=(0,0)$ the spin-spin correlation vanishes with $\vect{q}$, so we take QMC measurements at a small momentum deviation $\frac{2\pi}{L}$ away from $(0,0)$. Nevertheless, the momentum deviation $\vect{q}$ in the fitting formula \eqnref{eq:bessel} is still treated as a fitting parameter (of the order $\sim\frac{2\pi}{L}$) to partially take care of the finte-size effect. One can see that for the system sizes considered, $L=32,48,64$ and $96$ (the others are not shown), the Bessel functions in Eq.~\eqref{eq:bessel} fit the data well. In Fig.~\ref{fig:fig2} (a) and (b), we fit the imaginary time data with $\eta_{(\pi,0)}^{x}$ and $\eta_{(0,0)}^{z}$ as free fitting parameters.
Because the short (imaginary-)time data contain significant contributions from high energy excitations, for which the fitting function is no longer valid, we dynamically choose the fitting range starting from an appropriate short-time cutoff such that $\chi^{2}/d.o.f$ of the fitting is close to one. After fitting all system sizes from $L=16$ to $L=96$,  the scaling dimensions $\eta_{(\pi,0)}^{x}$ and $\eta_{(0,0)}^{z}$ are obtained, and their finite size scaling are given in Fig.~\ref{fig:fig3}.

The extrapolated values of the fitted scaling dimensions converge to zero for infinite size within numerical errors  as shown in Fig.~\ref{fig:fig3} . With the system size up to $L=96$ we obtain $\eta_{(\pi,0)}^{x}=0.002(9)$ and $\eta_{(0,0)}^{z}=0.004(6)$ indicating that the currents $J_{2}^{15}$ and $J_0^{12}$ are conserved. The conservation of $J_0^{12}$ is just the consequence of spin $U(1)$ symmetry, but the conservation of $J_2^{15}$ is a remarkable observation in favour of the emergent $\O(4)$ symmetry at the easy-plane DQCP.  


\begin{figure}[htp!]
\includegraphics[width=0.8\columnwidth]{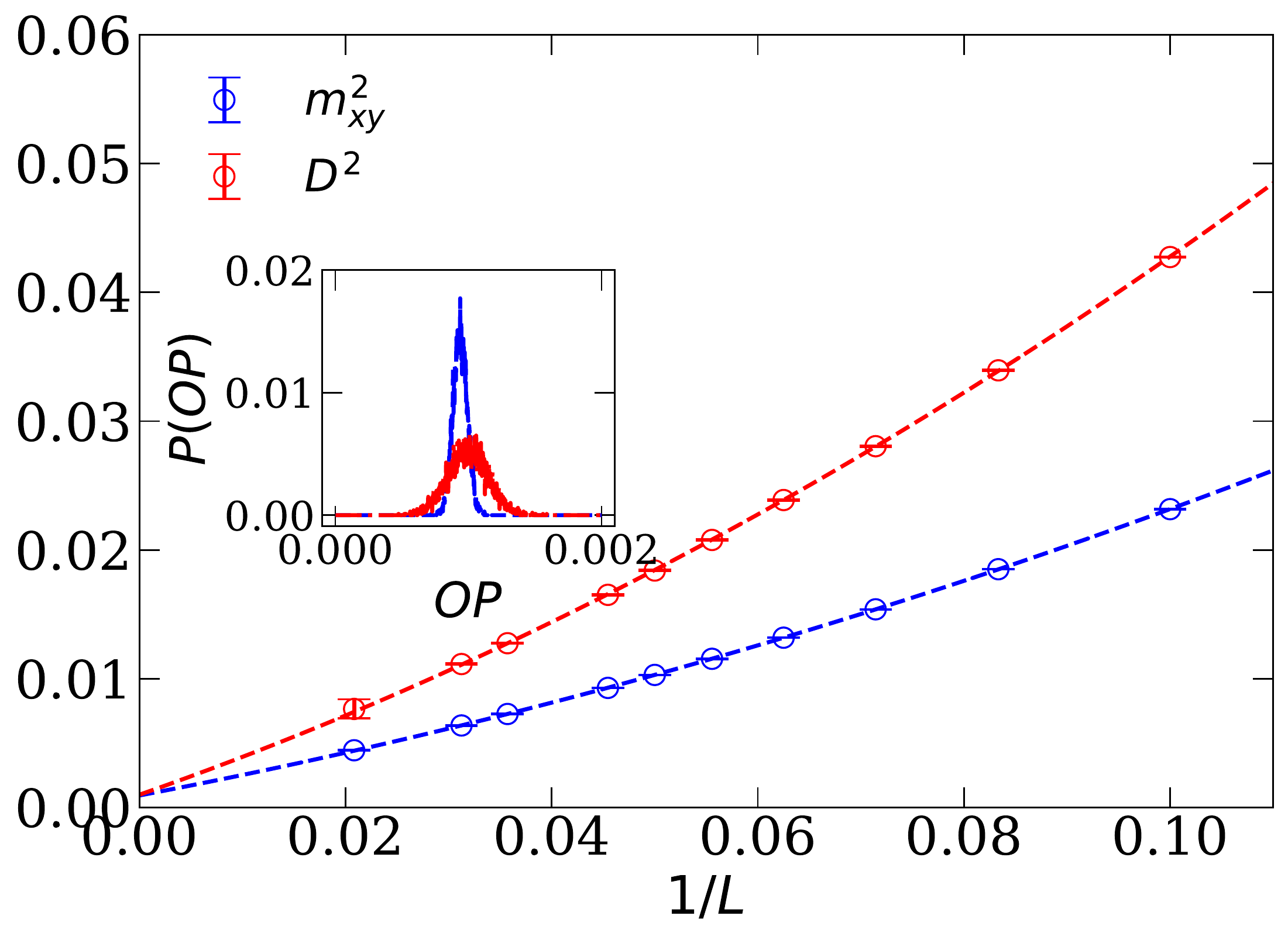}
\caption{The finite size extrapolation of the AFXY and VBS order parameters at the easy-plane DQCP. The critical points $q_c$ for each finite size $L$ are determined from their corresponding Binder ratio crossings, and the largest system size is $L=96$. Inset shows the histogram of extrapolated $\langle\vect{N}^2\rangle=0.0009(2)$ and $\langle\vect{D}^{2}\rangle=0.0010(3)$.}
\label{fig:fig4}
\end{figure}

{\it{Discussions.-}} Lastly, let's discuss the extrapolation of the order parameters at the DQCP. As shown in Fig.~\ref{fig:fig4}, the AFXY and VBS order parameters $\langle \vect{N}^{2} \rangle =\frac{1}{2}(\langle N^{2}_{x}+N^{2}_y \rangle)$ and $\langle \vect{D}^{2} \rangle =\frac{1}{2}(\langle D^{2}_{x}+D^{2}_y \rangle)$ are measured for various system sizes at their corresponding finite size $q_c(L)$ (the determination of $q_c(L)$ is discussed in Sec. III of SM~\cite{suppl}), and with system size up to $L=96$, the $1/L$ extrapolation gives very small (if not zero) $\langle \vect{N}^{2} \rangle = 0.0009(2)$ and $\langle \vect{D}^{2} \rangle = 0.0010(3)$ at the thermodynamic limit. So at this point, it is possible that the easy-plane DQCP is similar to the recently found symmetry-enhanced first-order transitions~\cite{BWZhao2018,Serna2018}, but with even weaker orders. It is also possible that both order parameters eventually flow to zero, in consistent with the original DQCP scenario, i.e.\,a continuous transition. We leave this to future studies.

To conclude, we have successfully demonstrated Noether's theorem in action in the frontier research of quantum matter -- to identify the emergent SO(4) continuous symmetry at easy-plane DQCP. Our attempt stands out as an new numerical tool in identifying emergent continuous symmetry, ubiquitously present at novel quantum phase transitions in DQCP, frustrated magnets, interacting topological phases and quantum electrodynamic systems. Comparing with the analyses of order parameter histogram and critical exponents, our approach provides a complementary view point both in numerical accessibility and theoretical elegance.

\begin{acknowledgments}
We thank Chong Wang, Yin-Chen He, Ashvin Vishwanath for helpful discussions and Anders Sandvik for insightful suggestions throughout the project. NSM and ZYM acknowledge the supports from the Ministry of Science and Technology of China through the National Key Research and Development Program (2016YFA0300502), the Strategic Priority Research Program  of the Chinese Academy of Sciences (XDB28000000), and the National Science Foundation of China (11574359). We thank the Centre for Quantum Simulation Sciences at Institute of Physics,  Chinese Academy of Sciences, and the Tianhe-1A platform at the National Supercomputer Centre in Tianjin for technical support and generous allocation of CPU time.
\end{acknowledgments}

\bibliographystyle{apsrev4-1}
\bibliography{current}

\setcounter{page}{1}
\setcounter{equation}{0}
\setcounter{figure}{0}
\renewcommand{\theequation}{S\arabic{equation}}
\renewcommand{\thefigure}{S\arabic{figure}}

\newpage

\begin{widetext}
\section*{Supplemental Material}

\centerline{\bf Emmy Noether looks at the deconfined quantum critical point}
	\vskip3mm

	\centerline{}

	\vskip6mm
In this supplemental material, we provide the theoretical description of the identification of conserved currents in the field theory (Sec.~SI), the technical details concerning the QMC measurements of conserved currents (Sec.~SII) and the finite size scaling of the order parameters at the easy-plane DQCP (Sec.~SIII).

\subsection{SI. Identification of currents in field theory}

The easy-plane DQCP can be described by the $N_f=4$ quantum electrodynamics (QED) theory with an easy-plane anisotropy.  In this description, the spin operator $\vect{S}_i$ is first fractionalized into two fermionic spinons $f_i=(f_{i\uparrow},f_{i\downarrow})^\intercal$ following $\vect{S}_i=\frac{1}{2}f_{i}^\dagger\vect{\sigma}f_{i}$, under the constraint $f_i^\dagger f_i=1$ on each site $i$. An emergent $\U(1)$ gauge structure (generated by the gauge transformation $f_{i}\to e^{\ii\theta_i}f_{i}$) arises from the fractionalization. Here the gauge group can be enlarged to $\SU(2)$, but this will not affect of our following discussion, so we will take a simpler $\U(1)$ gauge group instead. At the DQCP, the system can be viewed as a $\U(1)$ Dirac spin liquid, where the spinons are placed in a Dirac dispersion modeled by the square-lattice $\pi$-flux state. Specific spinon interactions will also be introduced to implement the easy-plane anisotropy. At the mean-field level (ignoring the gauge fluctuation and the spinon interaction for a moment), the spinon is described by $\pi$-flux model Hamiltonian
\begin{equation}\label{eq:HMF}
H_\text{MF}=\sum_{i}f_{i+\hat{x}}^\dagger f_{i}+(-)^{x_i} f_{i+\hat{y}}^\dagger f_{i}+\mathrm{h.c.}+\cdots,
\end{equation}
where each site $i$ is equivalently labeled by its coordinate $\vect{r}_i=(x_i,y_i)$ on the square lattice, and $\hat{x}=(1,0)$ and $\hat{y}=(0,1)$ are the lattice vectors. We take a $2\times2$ unit cell with the sublattices labeled by $A(0,0)$, $B(0,1)$, $C(1,0)$, $D(1,1)$ and define the spinon operator in the momentum space as $f_{\vect{k}}=(f_{\vect{k}A},f_{\vect{k}B},f_{\vect{k}C},f_{\vect{k}D})^\intercal$, where $f_{\vect{k}A}=\sum_{i\in A}f_{i}e^{-\ii\vect{k}\cdot\vect{r}_i}$ (and similar for other sublattices) and $\vect{k}\in[0,\pi)^{\times2}$. In the momentum space, the mean-field Hamiltonian $H_\text{MF}$ in \eqnref{eq:HMF} becomes
\begin{equation}
\label{eq:HMF k}
H_\text{MF}=2\sum_{\vect{k}}f_{\vect{k}}^\dagger (\cos k_x\sigma^{100}+\cos k_y \sigma^{310})f_{\vect{k}},
\end{equation}
where $\sigma^{\mu\nu\lambda}=\sigma^{\mu}\otimes\sigma^{\mu}\otimes\sigma^{\lambda}$ (with $\mu,\nu,\lambda=0,1,2,3$) denotes the direct product of Pauli matrices. The Hamiltonian in \eqnref{eq:HMF k} describes a spinon band structure with \emph{four} degenerated Dirac cones at the momentum $\vect{k}=(\pi/2,\pi/2)$, matching the matter content of the $N_f=4$ QED theory. In Ref.~\cite{Ma2018a}, it is explicitly shown that the spectral properties at the easy-plane DQCP can be qualitatively captured by the spinon mean-field theory in \eqnref{eq:HMF k} and gauge fluctuations and spinon interactions must be included to obtain quantitatively correct spin excitation spectra, but for the sake of identifying the emergent $\SO(4)$ current, we can ignore these interaction effects.

\begin{table}[htp!]
\begin{center}
\begin{tabular}{|c|c|c|}\hline
\multirow{2}{*}{$\vect{Q}$} & lattice model &  field theory\\ 
& $\vect{s}_\vect{Q}=$ & $\vect{S}_\vect{Q}\sim$ \\ \hline
$(0,0)$ & $(\sigma^{001},\sigma^{002},\sigma^{003})$ & $(J_0^{45},J_0^{53},J_0^{34})$\\
$(\pi,0)$ & $(\sigma^{301},\sigma^{302},\sigma^{303})$ & $(J_2^{23},J_2^{24},J_2^{25})$ \\
$(0,\pi)$ & $(\sigma^{031},\sigma^{032},\sigma^{033})$ & $(J_1^{13},J_1^{14},J_1^{15})$ \\
$(\pi,\pi)$ & $(\sigma^{331},\sigma^{332},\sigma^{333})$ & $(n^3,n^4,n^5)$ \\ \hline
\end{tabular}
\caption{Identification of the low-energy spin operators to the field theory operators.}
\label{tab:SQ}
\end{center}
\end{table}

It was found in  Ref.~\cite{Ma2018a} that gapless continuua appear in the spin excitation spectra at four distinct momenta $\vect{Q}$: $(0,0)$, $(\pi,0)$, $(0,\pi)$ and $(\pi,\pi)$. Spin operators at these momenta can be written in terms of spinon bilinear operators, 
\begin{equation}
\vect{S}_{\vect{Q}}=\sum_{i}\vect{S}_{i}e^{-\ii\vect{Q}\cdot\vect{r}_i}=\frac{1}{2}\sum_{\vect{k}}f_{\vect{k}}^\dagger \vect{s}_\vect{Q} f_{\vect{k}},
\end{equation}
where the matrix forms of $\vect{s}_\vect{Q}$ are listed in \tabref{tab:SQ} for various $\vect{Q}$. In particular, $\vect{S}_{(\pi,\pi)}$ corresponds to the N\'eel order parameter. The VBS order parameter $\vect{D}=\sum_{i}\vect{D}_i=\sum_{i}(D_i^x,D_i^y)$ can also be written in terms of spinon bilinear operators,
\begin{equation}
D_i^x= \tfrac{1}{2}(-)^{x_i}f_{i+\hat{x}}^\dagger f_{i}+\mathrm{h.c.},\quad D_i^y= \tfrac{1}{2}(-)^{x_i+y_i}f_{i+\hat{y}}^\dagger f_{i}+\mathrm{h.c.},
\end{equation}
which, in the momentum space, reads
\begin{equation}
\vect{D}=\sum_{\vect{k}}f_{\vect{k}}^\dagger (\sin k_x\sigma^{200},\sin k_y \sigma^{320})f_{\vect{k}}.
\end{equation}
We focus on the low-energy spinon $f$ and expand the spinon band structure in \eqnref{eq:HMF k} around the Dirac point at $\vect{k}=(\pi/2,\pi/2)$. The low-energy effective theory turns out to be a $N_f=4$ QED theory described by the following Lagrangian,
\begin{equation}
\label{eq:L QED}
\scL=\bar{f}\gamma^\mu(\partial_\mu -\ii a_\mu)f+\scL_\text{int}+\cdots,
\end{equation}
where $(\gamma^0,\gamma^1,\gamma^2)=(\sigma^{210},\sigma^{310},\sigma^{100})$ and $\bar{f}=f^\dagger\gamma^0$. $\scL_\text{int}$ contains the spinon interaction to be specified later. One can find the following five matrices $(\Gamma^1,\Gamma^2,\Gamma^3,\Gamma^4,\Gamma^5)=(\sigma^{010},\sigma^{130},\sigma^{121},\sigma^{122},\sigma^{123})$ in the spinon single-particle space, which anticommute with each other and all commute with the $\gamma^\mu$ matrices. They can be used to construct an $\O(5)$ vector $\vect{n}=\bar{f}\vect{\Gamma} f$ at the field theory level. The physical meaning of  the $\O(5)$ vector is a combined order parameter of the three-component N\'eel and two-component VBS order parameters,
\begin{equation}
\label{eq:orderparameters}
\begin{split}
(n^1,n^2)&=\bar{f}(\Gamma^1,\Gamma^2)f=f^\dagger (\sigma^{200},\sigma^{320})f\sim \vect{D},\\
(n^3,n^4,n^5)&=\bar{f}(\Gamma^3,\Gamma^4,\Gamma^5)f
=f^\dagger (\sigma^{331},\sigma^{332},\sigma^{333})f\sim \vect{S}_{(\pi,\pi)}.
\end{split}
\end{equation}
The $\SO(5)$ group that rotates the $\O(5)$ vector $\vect{n}$ is generated by ten generators $\Sigma^{a b}=\frac{1}{2\ii}[\Gamma^a,\Gamma^b]$ ($a,b=1,\cdots,5$). This $\SO(5)$ group is expected to be an emergent symmetry for the spin $\SU(2)$ symmetric DQCP. According to Noether's theorem, there exist ten emergent conserved currents associated to $\SO(5)$ symmetry generators as 
\begin{equation}
J_\mu^{ab}=\bar{f}\gamma^\mu\Sigma^{ab} f.
\end{equation}
However, in our case, the emergent $\SO(5)$ symmetry is broken down to $\O(4)$ explicitly by the easy-plane anisotropy, which singles out $n^5$ (N\'eel $z$-component) as a special component. This amounts to adding the interaction $\scL_\text{int}=\frac{g}{N_f\Lambda}(\bar{f}\Gamma^5 f)^3$ to the Lagrangian in \eqnref{eq:L QED} explicitly (where a momentum cutoff $\Lambda$ is introduced to render the coupling $g$ dimensionless). The remaining six $\SO(4)$ conserved currents are just a subset of $J_\mu^{ab}$ for $a,b=1,2,3,4$. With this setup, we can identify  the current operators $J_\mu^{ab}$ in the field theory to the spin operator $\vect{S}_{\vect{Q}}$ by matching their matrix representation $\vect{s}_\vect{Q}$ of spinon bilinear forms, as summarized in \tabref{tab:SQ}. Mathematically, the observation of \emph{all} the SO(4) conserved currents would imply the SO(4) symmetry. In our work we check five of them as $J_2^{23},J_{1}^{13},J_{2}^{24},J_{1}^{14},J_{0}^{34}$ in \tabref{tab:SQ}. And only the the first and the last one out of them were measured explicitly, but the microscopic $\mathrm{U}(1)\times\mathbb{Z}_4$ symmetry can relate the first four currents to each other, so it suffice to check just the first one among them,which is $J^{23}_2$.
It is associated to an emergent rotation symmetry between AFXY and VBS order parameters. Using the spinon representation, we found
\begin{equation}
J^{23}_2=\bar{f}\gamma^2\Sigma^{23}f=\frac{1}{2\ii}\bar{f}\sigma^{100}[\sigma^{130},\sigma^{121}]f=\frac{1}{2}f^\dagger\sigma^{301}f
\end{equation}
corresponds to the $S_{(\pi,0)}^x$ spin fluctuation (as $s_{(\pi,0)}^x=\sigma^{301}$), which are measured in QMC simulation. For comparison, we also study another conserved current $J_0^{34}$ in association with the microscopic $\U(1)$ spin rotation symmetry, which appears as the $S_{(0,0)}^z$ spin fluctuation following the similar derivation.

The only current that we missed is associated to the rotation between VBS order parameters, as it does not manifest itself in the two-point spin correlation (only detectable in higher order correlations). However due to the non-Abelian nature of the SO(4) group, this last generator can be inferred from the first five generators, so the conserved result of the other five current operators does also provide an \emph{indirect} evidence for the last conservation current. Besides, the $\mathbb{Z}_2$ subgroup  of the full O(4) symmetry does not have a corresponding conserved current as it is not continuous. But it  can be argued from the microscopic $\mathbb{Z}_2$ spin flip symmetry of $(S^x,S^y,S^z)\to(-S^x,S^y,-S^z)$. Therefore, our results indeed provide sufficient evidences to support the emergent O(4) symmetry at the easy-plane DQCP.

\subsection{SII. Conserved and non-conserved currents}

In this section, we will give a detailed derivation of the fitting function we proposed in \eqnref{eq:bessel} in the main text. Let us start with the more general form of the current-current correlation with an anomalous exponent $\eta$,
\eq{\langle J_\mu^{ab}(-q)J_\nu^{ab}(q)\rangle\sim|q|^{1+\eta}\Big(\delta_{\mu\nu}-\frac{q_\mu q_\nu}{|q|^2}\Big)=\frac{|q|^2\delta_{\mu\nu}-q_\mu q_\nu}{|q|^{1-\eta}}}
without the assumption that the current is conserved. The result can later be applied to the case of conserved current by setting $\eta=0$.

To make connection to the imaginary time data measured in the QMC simulation, we need to Fourier transform $\langle J_\mu^{ab}(-q)J_\nu^{ab}(q)\rangle$ from the imaginary frequency $q_0$ to the imaginary time $\tau$ domain (where $q=(q_0,\vect{q})$ contains both the imaginary frequency and the momentum components),
\eq{\Pi_{\mu\nu}(\tau,\vect{q})=\int\dd q_0 e^{-\ii q_0 \tau}\langle J_\mu^{ab}(-q)J_\nu^{ab}(q)\rangle.}
To carry out the Fourier transform, we use the mathematical fact that
\eqs{&\int\dd q_0 e^{-\ii q_0 \tau}\frac{1}{|q|^{1-\eta}}=\int\dd q_0 e^{-\ii q_0 \tau}\frac{1}{(q_0^2+\vect{q}^2)^{(1-\eta)/2}}\\
&=\frac{2\sqrt{\pi}}{\Gamma(\frac{1-\eta}{2})}\Big|\frac{2\vect{q}}{\tau}\Big|^{\eta/2}K_{\frac{\eta}{2}}(|\vect{q}\tau|)\\
&=\frac{2\sqrt{\pi}}{\Gamma(\frac{1-\eta}{2})}F_{\frac{\eta}{2}}(\tau,\vect{q}),\label{eq:Fouriermath}}
where the function $F_\alpha(\tau,\vect{q})=|\frac{2\vect{q}}{\tau}|^\alpha K_\alpha(|\vect{q}\tau|)$ was introduced to simplify the notation. With this we can evaluate the following
\eqs{\Pi_{00}(\tau,\vect{q})&=\int\dd q_0 e^{-\ii q_0\tau}\frac{\vect{q}^2}{|q|^{1-\eta}}\\
&=\frac{2\sqrt{\pi}}{\Gamma(\frac{1-\eta}{2})} \vect{q}^2 F_{\frac{\eta}{2}}(\tau,\vect{q}),}
\eqs{\Pi_{22}(\tau,\vect{q})&=\int\dd q_0 e^{-\ii q_0\tau}\frac{|q|^2-q_2^2}{|q|^{1-\eta}}\\
&=\int\dd q_0 e^{-\ii q_0\tau}\Big(\frac{1}{|q|^{1-(\eta+2)}}-\frac{q_2^2}{|q|^{1-\eta}}\Big)\\
&=-\frac{2\sqrt{\pi}}{\Gamma(\frac{1-\eta}{2})} \Big(\tfrac{\eta+1}{2}F_{\frac{\eta}{2}+1}(\tau,\vect{q})+q_2^2 F_{\frac{\eta}{2}}(\tau,\vect{q})\Big).}
We will use these results later.

The current-current correlations are measured as spin-spin correlations around different momenta in different spin channels. We will focus on the following two correlations
\eqs{G_{(0,0)}^{a}(\tau,\vect{q})&=\int\dd q_0 e^{-\ii q_0 \tau} \langle S_{(0,0)}^a(-q)S_{(0,0)}^a(q)\rangle.\\
G_{(\pi,0)}^{a}(\tau,\vect{q})&=\int\dd q_0 e^{-\ii q_0 \tau} \langle S_{(\pi,0)}^a(-q)S_{(\pi,0)}^a(q)\rangle.}
According to \tabref{tab:SQ} in the main text, the operator $S_{(0,0)}^a$ corresponds to the current $J_{0}^{bc}$ and the operator $S_{(\pi,0)}^a$ corresponds to the current $J_{2}^{bc}$. So we can make the following identifications
\eqs{G_{(0,0)}^{a}&\sim\Pi_{00}\propto \vect{q}^2F_{\frac{\eta}{2}}(\tau,\vect{q}),\\
 G_{(\pi,0)}^{a}&\sim\Pi_{22}\propto  q_2^2F_{\frac{\eta}{2}}(\tau,\vect{q})+\tfrac{\eta+1}{2}F_{\frac{\eta}{2}+1}(\tau,\vect{q}),}
which lead to the fitting functions in \eqnref{eq:bessel} in the main text.

For conserved currents, we expect $\eta=0$, thus
\begin{equation}
\begin{split}
G^z_{(0,0)}(\tau,\vect{q})&\propto\vect{q}^2K_0(|\vect{q}\tau|),\\
G^x_{(\pi,0)}(\tau,\vect{q})&\propto q_2^2K_0(|\vect{q}\tau|)+|\vect{q}/\tau|K_{1}(|\vect{q}\tau|).
\end{split}
\end{equation}
Around $\vect{Q}=(0,0)$ the spin-spin correlation vanishes with $\vect{q}$, so the measurement can only be done at a small momentum deviation $\vect{q}$ away from $(0,0)$. Around $\vect{Q}=(\pi,0)$, the spin-spin correlation remains finite as $\vect{q}\to0$, which takes the simple power-law form
\begin{equation}
G^x_{(\pi,0)}(\tau,\vect{q}=0)\propto |\tau|^{-2}.
\end{equation} 

\subsection{SIII. Extrapolation of the order parameters}
\label{sec:order}
In this section, we provide details of the determination of the position $q_c(L)$ of the DQCP of EPJQ model, and explain how the extrapolation of the order parameters are performed in  Fig.~\ref{fig:fig4} in the main text.

\begin{figure}[htp!]
\includegraphics[width=0.5\columnwidth]{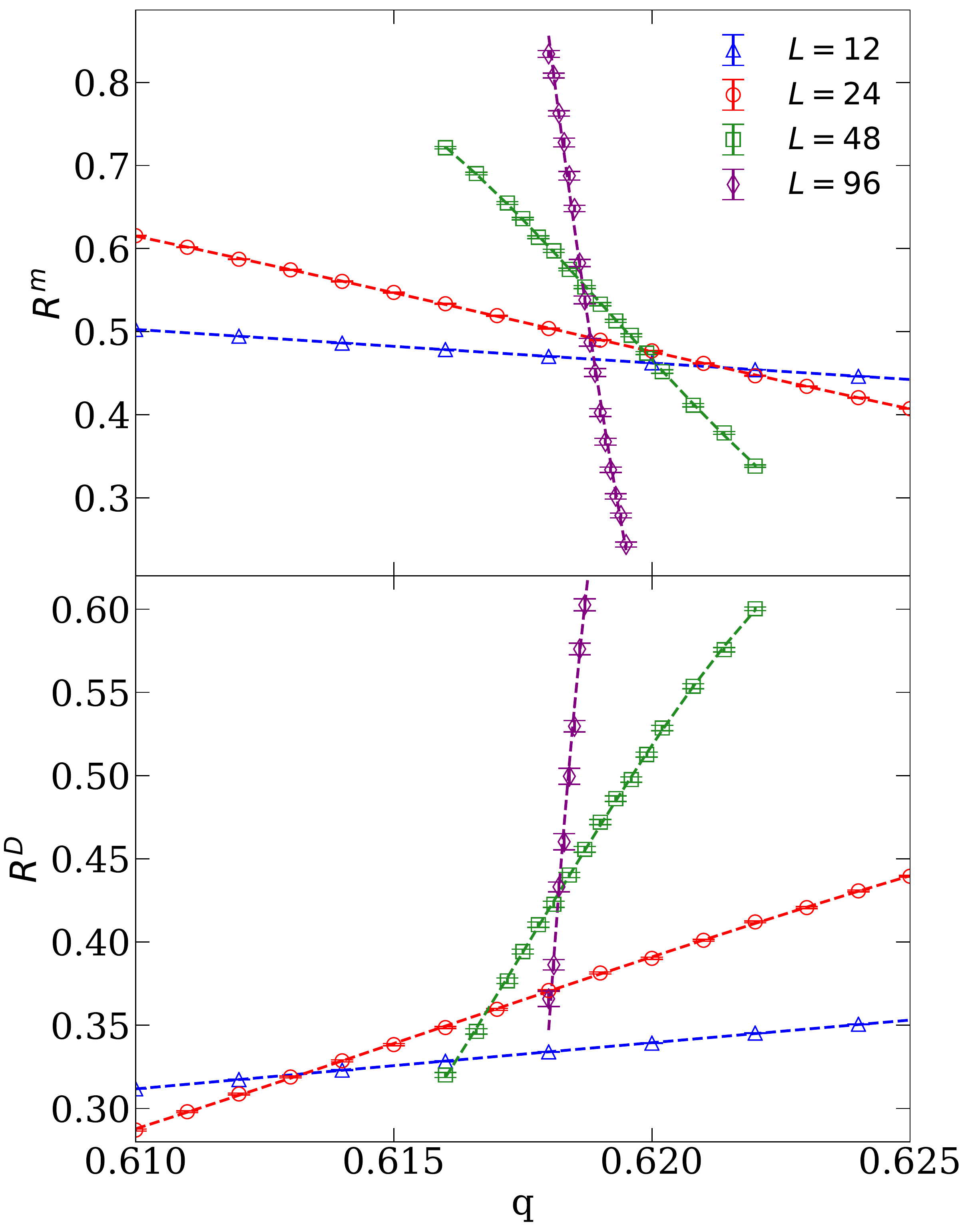}
\caption{Determination of the transition point between AFXY and VBS phases in the EPJQ model in Eq.~\eqref{eq:EPJQ} in main text. From the Binder ratios of $R^{m}$ and $R^{D}$ in Eq.~\eqref{eq:defineR}, the crossings of two consecutive $L$-s give $q_{c}(L)$.}
\label{fig:fig_app_1}
\end{figure}

\begin{figure}[htp!]
\includegraphics[width=0.5\columnwidth]{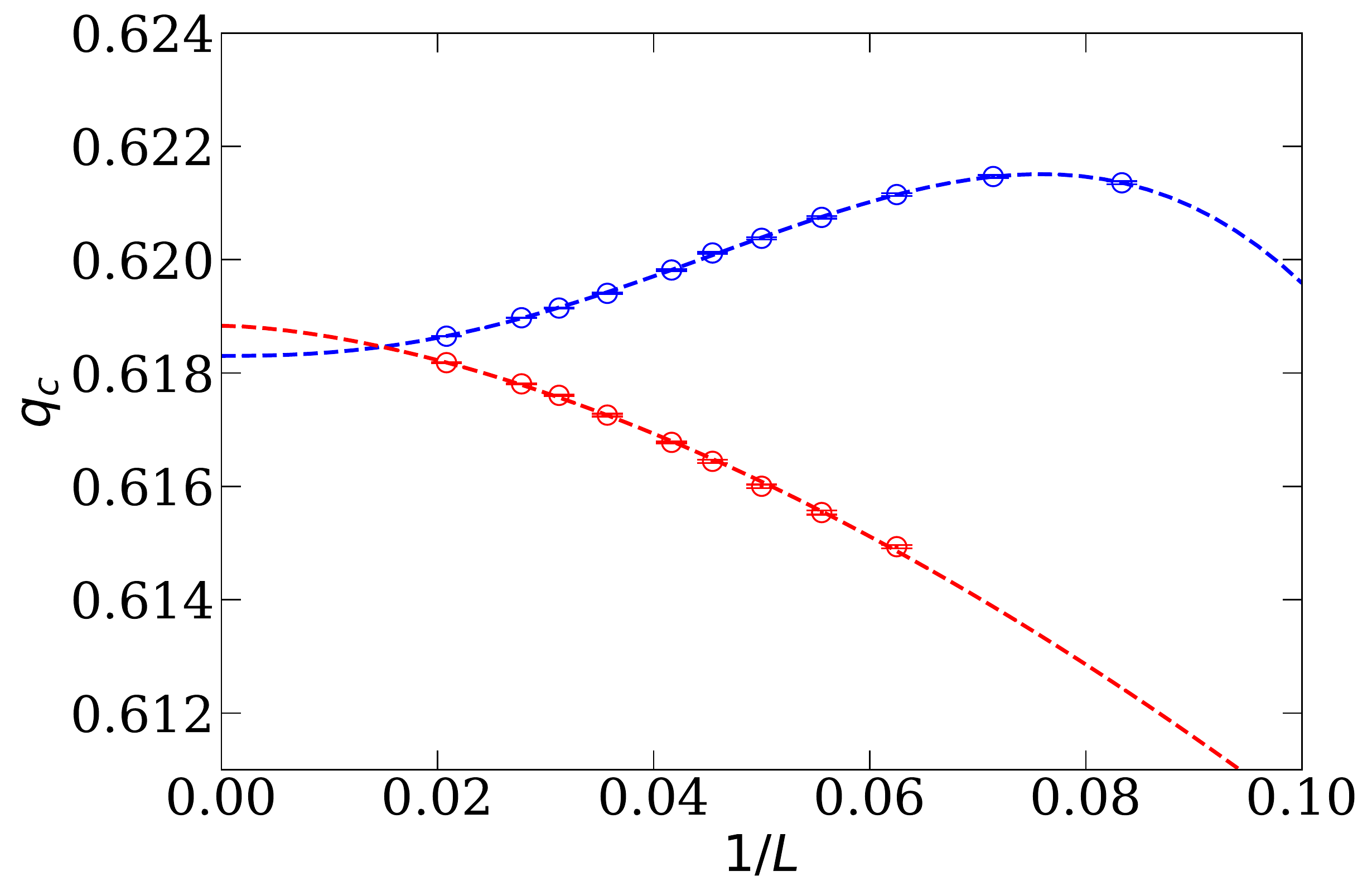}
\caption{Extrapolation of the $q_c$ obtained from Fig.~\ref{fig:fig_app_1} verse $1/L$. The blue circles are crossing values obtained from $R^{m}$ for different $L$ while red ones obtained from $R^{D}$.The dashed curves are the fitting functions of all data shown here using 
Eq.~\eqref{eq:ffsizescaling}. Thermodynamic limit values of $q_{c}(\infty)$ can be obtained as the extrapolated value at $1/L=0$.}
\label{fig:fig_app_2}
\end{figure}

First we measured the Binder ratio for both AFXY order parameter and VBS order parameters, given as

\begin{eqnarray}
R^{m} &=& \frac{\langle m_{xy}^{4} \rangle}{\langle m_{xy}^{2}\rangle^{2}}\nonumber\\
R^{D} &=& \frac{\langle D^{4}\rangle}{\langle D^{2} \rangle^2} 
\label{eq:defineR}
\end{eqnarray}
where $m_{xy}^{2} = \frac{1}{2}(m^{2}_{x}+m^{2}_{y})$ with $m^{2}_{x}=(\frac{1}{L^2}\sum_{i}(-1)^{i_x+i_y}S^{x}_{i})^2$ the square of magnetic moment along the $S^{x}$ with $(\pi,\pi)$ order on the lattice. And $D^{2}=\frac{1}{2}(D^{2}_{x}+D^{2}_{y})$ with $D^{2}_{x}=(\frac{1}{L^2}\sum_{i}(-1)^{i_x}\mathbf{S}_{i}\cdot\mathbf{S}_{i+x})^{2}$ the square of the dimer singlet along the $x$-axis with the $(\pi,0)$ order on the lattice.

Fig.~\ref{fig:fig_app_1} shows the crossing of the two Binder ratios at some representative system sizes. As $L$ increases, the drift of the Binder ratio can be seen, both in $R^{m}$ and $R^{D}$. From the crossing of two consecutive $L$-s, for example $L$ and $2L$, the finite size transition point $q_c(L)$ can be determined. And the size dependence of $q_{c}(L)$ is given by 
\begin{equation}
q_{c}(L)=q_{c}(\infty)+aL^{-1/\nu-\omega_{1}}+...
\label{eq:ffsizescaling}
\end{equation}
where $q_{c}(\infty)$ is the transition point at the thermodynamic limit and $\nu$ is the correlation length exponent and $\omega_1$ is the correction exponent. We then fit the crossing points $q_c(L)$ from both $R^{m}$ and $R^{D}$ using 
Eq.~\eqref{eq:ffsizescaling} and extrapolate to $q_{c}(\infty)$. In Fig.~\ref{fig:fig_app_2} we find the anisotropic behaviour of $q_c(L)$ from $R^{m}$, so one more term $a_{2}L^{-1/\nu-\omega_{2}}$ in Eq.~\eqref{eq:ffsizescaling} is included in the fitting
~\cite{nvsenma2018} to take care of such higher order corrections. At the thermodynamic limit the two order parameters actually extrapolate to the same value within errorbar as $q_{c}=0.61832(8)$ from $R^{m}$ and $q_{c}=0.6187(5)$ from $R^{D}$.
In Fig.~\ref{fig:fig4} of the main text, we plot $m^{2}_{xy}$ and $D^2$ for each system size $L$ at their corresponding $q_{c}(L)$, obtained in Fig.~\ref{fig:fig_app_1} here, and then as $L\to\infty$, the value of the order parameters at the thermodynamic limit, with $L=12,24,36,48,\cdots,84,96$, are obtained.

\end{widetext}


\end{document}